\documentclass[%
 reprint,
]{revtex4-2}

\usepackage{graphicx}
\usepackage{dcolumn}
\usepackage{bm}

\begin{document}

\title{Ternary fission with the emission of long-range $\alpha$ particles in fission of the heaviest nuclei}

\author{J. Khuyagbaatar} \email{J.Khuyagbaatar@gsi.de}
\affiliation{GSI Helmholtzzentrum f\"ur Schwerionenforschung, 64291 Darmstadt, Germany}

\date{\today}

\begin{abstract}

The most probable outcome of the ternary fission is the emission of two heavy fragments and one light-charged particle. In about 90\%, these are $\alpha$ particles, which often referred to as the long-range alpha (LRA). Such decay has been extensively studied over decades in various heavy fissioning systems. The probability of such a process has been found to be about (2-4)$\times 10^{-3}$ relative to binary fission. The experimental data showed an increasing trend in the probability of such a process with an increase in fissility parameter within the range of $35–39$. In the last decades, a region of the heaviest nuclei has been substantially expanded in both proton and neutron numbers. This includes neutron-deficient heavy and superheavy nuclei with fissility parameters, which are significantly exceeding the aforementioned range.

In the present work, the currently available experimental data on the probability of the LRA emission, which is a representative of ternary fission, are discussed. The probabilities of LRA emission were calculated within the various empirical approaches, including the presently suggested semi-empirical expression. The latter one was derived on the basis of the $\alpha$-decay property of the fissioning nucleus. The results of all approaches discussed show that the probabilities of LRA emission are substantial (up to a percentage) in the fission of neutron-deficient heavy and superheavy nuclei.

\end{abstract}


\maketitle


\section{Introduction}
Nuclear fission primarily results in the division of two nuclei (fragments) with medium charges and masses. Since its discovery in 1939 \cite{Hahn39}, the fission phenomenon has been extensively investigated. As a consequence, many features of the fission process have been experimentally established and theoretically understood (see, e.g., Refs.~\cite{Flerov40,VanH73,Andreyev_2017,Fritz17}).

In the mid-1940s, several research groups reported the discovery of ternary (and also quaternary) fission \cite{Tsien47a,GREEN1947,SAN-TSIANG1947,Farwell47}. In these experimental studies, identifications of fission fragments (FFs) were based on their stopping ranges in a medium.
Accordingly, their discoveries were based on the observation of long tracks that were accompanied by two shorter ones, which correspond to the FFs. Such a long-tracked particle has been identified as the $^4$He nucleus, i.e., the $\alpha$ particle with an average energy of around 16~MeV. Consequently, such $\alpha$ particles have been referred to as long-range alpha (LRA), which historically was used for the high-energy $\alpha$ particles emitted from excited states of the nucleus \cite{Ruth16}.

Since then, many experimental studies of the LRA emission from fissions of different heavy nuclei have been conducted. It turned out that LRA emission is only one outcome of ternary fission, in which the charge and mass of the third fragment can become as large as those of two binary fragments \cite{Halpern71}. However, the probability of the LRA emission is about 90\% of all ternary fission, and it occurs once in every 300–400 cases of binary fission.

The probability of LRA particle emission is defined relative to the total amount of the binary fission.

\begin{equation}\label{PLRA_def}
	P_{\rm{LRA}} = N_{\rm{F1F2\alpha}} / N_{\rm{F1F2}},
\end{equation}

where $N_{\rm{F1F2}}$ and $N_{\rm{F1F2\alpha}}$ are the numbers of fission events detected with only binary fragments and with a third fragment as the LRA.

Studies on LRA emission have been conducted in spontaneously fissioning nuclei and in fission of excited nuclei that are produced in photon-, neutron-, proton-, $^4$He- and heavy-ion induced reactions (see, e.g., in Refs.~\cite{Coleman64,Loveland67,GUET19791,WAGEMANS19811,Wild85,Wagemans86,Verboven94,Ivanov1996,SEROT199834,VERMOTE20081,Thesis_Vermote,VERMOTE2010176} and references therein).

\begin{figure}[b]
	\vspace*{-3mm}
	\centering
	\hspace*{-2mm}
	\resizebox{0.48\textwidth}{!}{\includegraphics[angle=0]{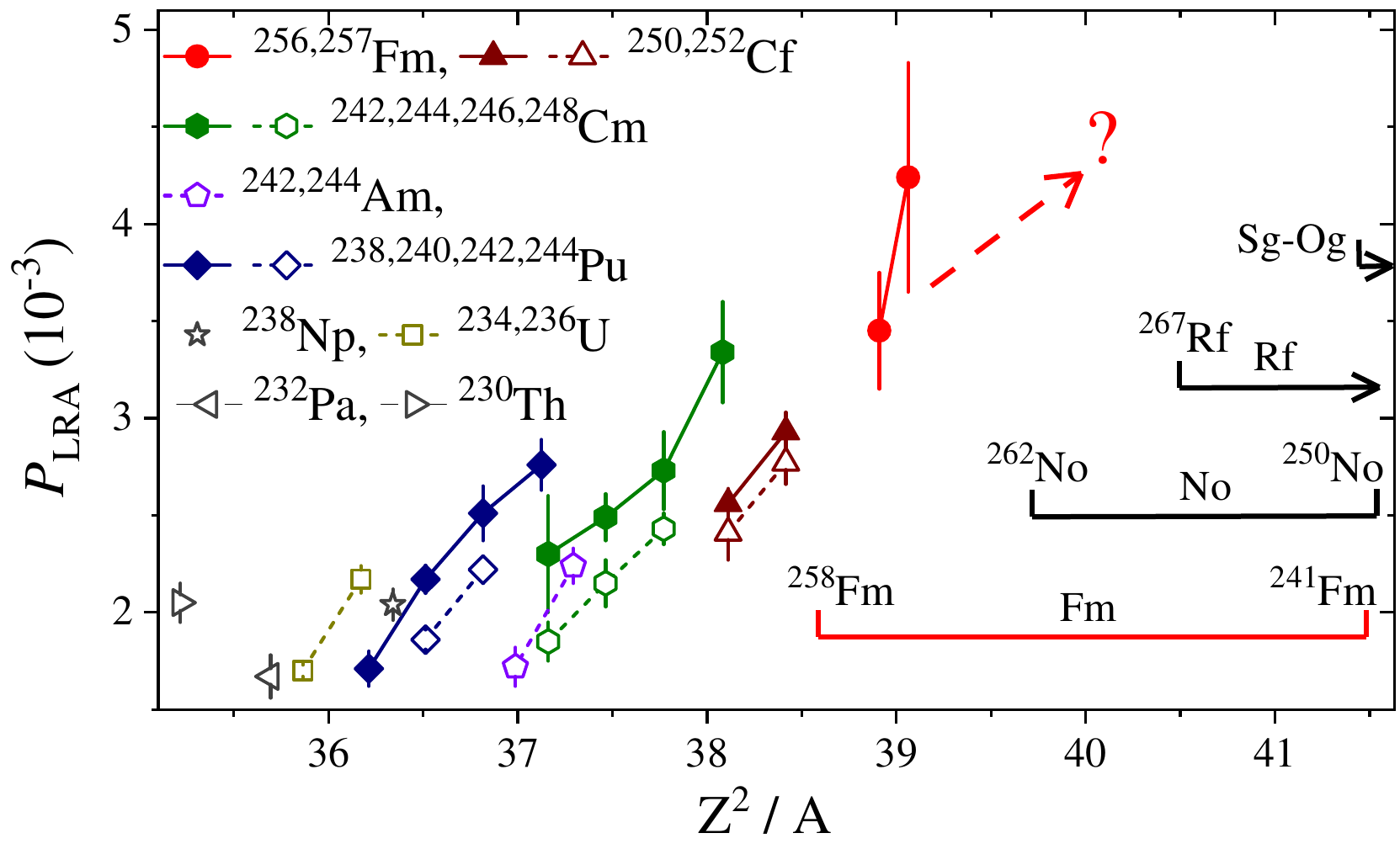}}
	\vspace*{-3mm}
	\caption{(color online) Experimental $P_{\rm{LRA}}$ values measured in the spontaneous and neutron-induced fissions marked by full and open symbols, respectively, are plotted as a function of the $Z^2/A$ of the fissioning nucleus. Isotopes corresponding to each symbol are given. The $P_{\rm{LRA}}$ values of isotopes of the same element are connected by the full and dotted lines. The experimental data are taken from the Refs.~\cite{WAGEMANS19811,Wild85,Wagemans86,SEROT199834,Ivanov1996,VERMOTE20081,Thesis_Vermote,VERMOTE2010176}. Regions for $Z^2/A$ values of the known isotopes of Fm-Og elements are marked.}
	\label{PLRA_Z^2/A_exp}
\end{figure}

The radioactive heavy isotopes of actinide elements, which can be produced in macroscopic amounts at dedicated facilities \cite{RobA15a}, have mostly been used as targets. Accordingly, to some extent, the range of fissioning nuclei for LRA emission studies was constrained in terms of $Z$ and $A$. The experimental data measured at higher excitation energies (e.g., $>$12~MeV), however, could be effected by the complex de-excitation process of the compound nucleus such as a multi-chance fission \cite{Khu15b,Hiro17}.

The most comprehensively studied cases of LRA emissions are $^{238,240,242,244}$Pu \cite{Wagemans86,SEROT199834}, $^{242,244,246,248}$Cm \cite{Ivanov1996,VERMOTE20081}, $^{250,252}$Cf \cite{Fraenkel67,Loveland74,Ivanov1996,VERMOTE2010176,RAMAYYA2001221,OBERSTEDT2005173} and $^{256,257}$Fm \cite{Wild85}. These data, acquired from both spontaneous and thermal-neutron-induced fission, where the excitation energy does not exceed $\approx$7~MeV, represent the most recent and valuable experimental source for elucidating the empirical features of the LRA emission process. It is noteworthy to mention that the measurement of LRA is a very challenging topic that requires solutions for various technical aspects, such as the safe handling of highly radioactive materials over extended periods, which may easily be longer than a few weeks. For instance, five months were spent for the measurement of 1321 LRA particles from the spontaneous fission (SF) of $^{240}$Pu \cite{SEROT199834}.

All experimental $P_{\rm{LRA}}$ values measured in heavy nuclei including the aforementioned cases, are shown in Fig.~\ref{PLRA_Z^2/A_exp} as a function of the fissility parameter, $Z^2/A$. Most of the data points were taken from ~Refs.~\cite{WAGEMANS19811,Wild85,Wagemans86,SEROT199834,Ivanov1996,VERMOTE20081,Thesis_Vermote,VERMOTE2010176}, where available experimental data on LRA have been evaluated.  The experimental data from the thermal neutron-induced reactions show about a 20\% reduction in $P_{\rm{LRA}}$ when the nucleus is excited (see Fig.~\ref{PLRA_Z^2/A_exp} and, e.g., Refs.~\cite{SEROT199834,VERMOTE20081}). Both types of experimental data show a linear-like dependence between the $P_{\rm{LRA}}$ and $Z^2/A$ (up to $Z^2/A \approx$39). Such a feature, noticed decades ago \cite{Nobles62}, is often used to interpret and discuss the LRA emission process \cite{Rubchenya88}. This link was explained within the liquid drop model, where the fission is described as the interplay between the Coulomb and the surface-tension forces, thus expressed in $Z^2/A$ \cite{BohW39}. Meanwhile, the observed large energy of LRA is often attributed to being formed at the cost of deformation energy \cite{Halpern71}. This led to the extension of the $P_{\rm{LRA}}\sim Z^2/A$ correlation further, where the deformation energy is expressed as a fission-energy available for the dissipation to the internal degrees of freedom, i.e., the difference between the $Q$-value and average total kinetic energy of fission \cite{Halpern71,Wild85}. 

Meanwhile, to date, the heaviest nuclei with $Z$ up to 118 are known. In many cases, their SF branches have been observed \cite{Fritz17,Khu20a,Oga22c}. As a result, SF is known in nuclei within the significantly wide range of $Z^2/A$ \cite{Khu20a}. As an example, one of the six known isotopes of element Fl~($Z=114$) with the mass number $A=286$, which decays by both $\alpha$-particle emission and SF \cite{nndc}, exhibits a $Z^2/A$ value of $\approx$45.4. This is significantly larger than the maximum value ($\approx$39) of the nucleus, where the LRA emission was observed (see Fig.~\ref{PLRA_Z^2/A_exp}). To emphasize the significance of the LRA emission in the fission of superheavy nuclei (SHN), the ranges of the $Z^2/A$ values of some known isotopes of the element Fm-Og are shown in Fig.~\ref{PLRA_Z^2/A_exp}. It is worth noting that larger $Z^2/A$ values correspond to neutron-deficient heavy nuclei and SHN, where SF is their predominant decay mode. According to the trend of the experimental data shown in Fig.~\ref{PLRA_Z^2/A_exp}, significantly large probabilities of LRA emission are anticipated in both known neutron-deficient heavy nuclei and SHN. 

Many works on the theoretical description of the ternary fission exist \cite{Halpern71,Carjan75,Carjan76,Serot2000,Andreev_2006,Zagrev10,Denisov02,Rubchenya88,VONOERTZEN2015223,Ren22}. 
However, most of these studies primarily focus on a detailed description such as dynamics of process and emissions of other charged particles of the ternary fission. In addition, the main interest of theoretical efforts are concerning the ternary fission into fragments with comparable masses. Therefore, presently, both experimental and theoretical studies on the LRA emission in the fission of the heaviest nuclei are not under active consideration. 

Therefore, the aim of this work is to emphasize the significance of the LRA emission in the fission of the heaviest nuclei and to renew interest in it, especially experimental. I will consider different approaches as alternatives for estimating the probability of the LRA emission in the heaviest nuclei.

\section{Experimental data and empirical features of LRA emission from the spontaneous fission}

There are several methods for predicting $P_{\rm{LRA}}$ in the fission of heavy nuclei. In the following, I will consider some of them, focusing only on the SF data. 

\subsection{Estimation of $P_{\rm{LRA}}$as a function of fissility parameter}

\begin{figure}	
	\vspace*{0mm}
	\centering
	\hspace*{0mm}
	\resizebox{0.44\textwidth}{!}{\includegraphics[angle=0]{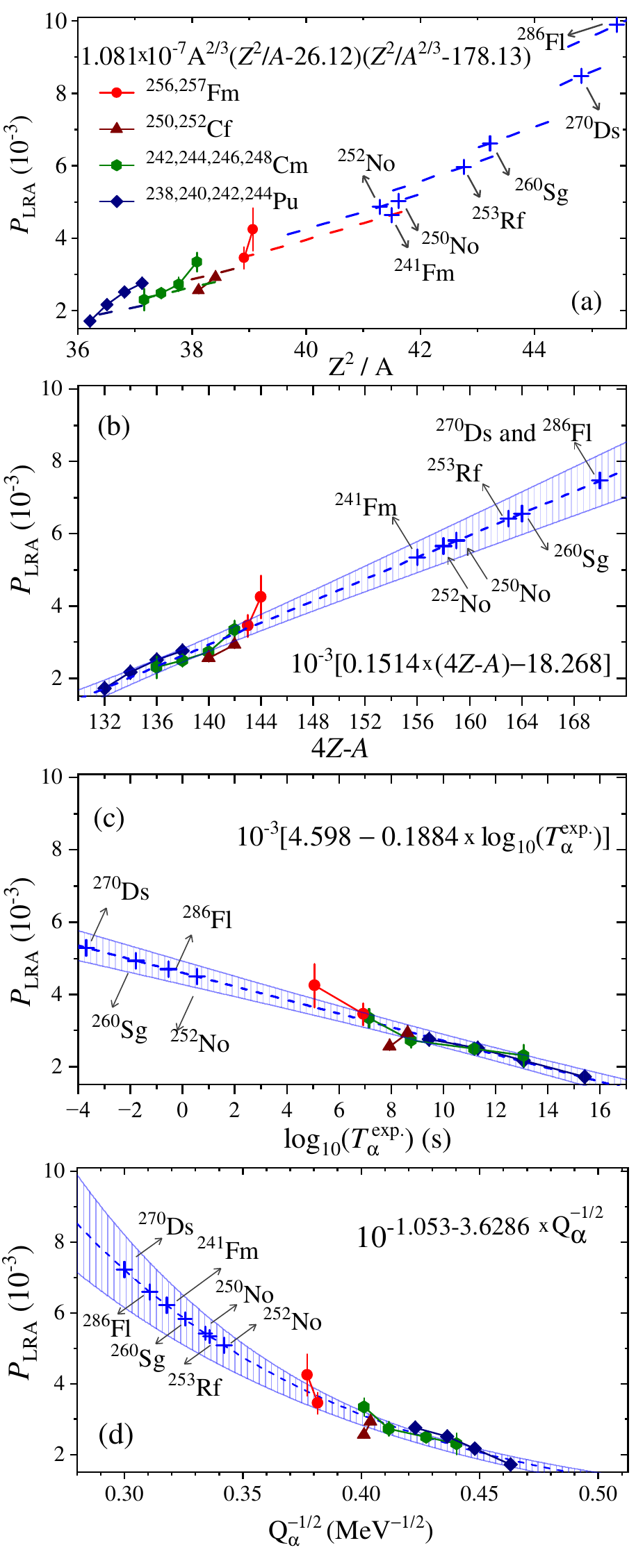}}
	\vspace*{-5mm}
	\caption{(color online) $P_{\rm{LRA}}$ measured in the cases of the spontaneous fissioning nuclei are shown as functions of (a) $Z^2/A$, (b) $4Z-A$, (c) ${\rm log}_{10}(T^{\rm{exp.}}_{\alpha})$, and (d) $Q^{-1/2}_{\alpha}$ variables. The experimental data shown by symbols. The full lines connect the $P_{\rm{LRA}}$ values of isotopes of the same element. (a) The dashed lines show the predictions for isotopes of the particular element according to the formula (given in the panel) from Ref.~\cite{Rubchenya88}. The expressions and dashed lines shown in (b)-(d) correspond to the fit of the experimental $P_{\rm{LRA}}$. Fit functions in (b), (c), and (d) are according to Refs.~\cite{Halpern71}, \cite{Wagemans86} and the present work, respectively. The sparse regions show boundaries of fits within the 68\% prediction level. Predicted $P_{\rm{LRA}}$ values for the several known heavy and superheavy nuclei are highlighted. See the text for details.
}
	\label{P_LRA_empirical}
\end{figure}

In the previous section, one of the empirical correlations, where the $P_{\rm{LRA}}$ was expressed as a function of $Z^2/A$, was already mentioned. This dependence has been formulated in the theoretical description of ternary fission within the dynamical approach by Rubchenya and Yavshits in Ref.~\cite{Rubchenya88}. In Fig.~\ref{P_LRA_empirical}(a), the results of their expression, which are downscaled by a factor of 0.9, are shown. This scaling factor was used to match the calculated total probability of the ternary fission by Rubchenya-Yavshits with the experimental $P_{\rm{LRA}}$ values from the SF cases, which are also given in Fig.~\ref{P_LRA_empirical}(a). As one can see, this expression, which has a direct dependence on the $Z^2/A$ variable, provides a reasonable description of the experimental data and their tendency. Apart from this, however, in particular cases, agreements between predictions and experiments are poor. By comparing the experimental and predicted values, one can find out that this approach describes the experimental data within about 15\% of uncertainty.  According to this prediction, each isotopic chain of the particular element has a linear dependence as a function of $Z^2/A$. This indicates that the LRA emission is more favorable for the most neutron-deficient isotopes of the particular element. Meanwhile, $P_{\rm{LRA}}$ has a much stronger, i.e., a quadratic dependence on $Z$, which should favor the LRA emission in heavier $Z$ nuclei. In order to highlight these features, the predicted $P_{\rm{LRA}}$ values for some of the known heaviest nuclei are marked. 
As expected, in the fission of SHN, the LRA emission becomes a significant process, which can be seen in the cases of $^{270}$Ds and $^{286}$Fl, where probabilities are reaching 1\%. 

\subsection{Estimation of $P_{\rm{LRA}}$ as a function of proton and neutron numbers}

Empirically observed a linear correlation between the $P_{\rm{LRA}}$ and $4Z-A$ variables was actually one of the first approaches to estimating the $P_{\rm{LRA}}$ \cite{Coleman64,Loveland67,Halpern71}. The experimental $P_{\rm{LRA}}$ values from the SF cases are shown in Fig.~\ref{P_LRA_empirical}(b) as a function of $4Z-A$. They do, in fact, follow a straight line. The result of fit is also shown in Fig.~\ref{P_LRA_empirical}(b). It predicts a further increase in $P_{\rm{LRA}}$ with an increase in $Z$. On the other hand, for isotopes of a particular element, it predicts a decrease of $P_{\rm{LRA}}$ with an increase in $A$, which is in agreement with the experimental observation. As an example, the $P_{\rm{LRA}}=2.76(13)\times10^{-3}$ for $^{238}$Pu is about 1.6 times larger than the $P_{\rm{LRA}}=1.71(9)\times10^{-3}$ of the six neutron heavier $^{244}$Pu isotope \cite{SEROT199834}. This approach is consistent with the above-discussed $P_{\rm{LRA}}\sim Z^2/A$ correlation, where the LRA emission is favorable in more neutron-deficient heavy nuclei.

The estimated $P_{\rm{LRA}}$ for several neutron-deficient heavy nuclei and SHN, similar to the above discussion, are shown in Fig.~\ref{P_LRA_empirical}(b). Again, these values are significantly larger than those for the known cases of the LRA emission. Eventually, the same $P_{\rm{LRA}}$ values were predicted for $^{270}$Ds and $^{286}$Fl, which have the same $4Z-A=170$. Accordingly, this indicates the limitation of the $4Z-A$ systematics.

\subsection{Estimation of $P_{\rm{LRA}}$ as a function of alpha-decay half-life}

Another interesting feature of the LRA emission was noticed when the $P_{\rm{LRA}}$ values are presented as a function of $-\rm log_{10}(\lambda_{\alpha})$ \cite{Wagemans86}. $\lambda_{\alpha}$ is a radioactive $\alpha$-decay constant for the ground-state of the fissioning nucleus. In Fig.~\ref{P_LRA_empirical}(c), such a dependence is shown but as a function of ${\rm log}_{10}(T^{\rm{exp.}}_{\alpha})$. $T^{\rm{exp.}}_{\alpha}$ is the experimental partial $\alpha$-decay half-lives of the ground state of the fissioning nucleus \cite{nndc}.
As one can see, the observed correlation shows a well-pronounced linear dependency, which was fitted as shown in Fig.~\ref{P_LRA_empirical}(c). However, the implication of such a fit-result have a strict limit since the $T^{\rm{exp.}}_{\alpha}$ is not always known for the ground state. Especially, this concerns the odd-$A$ and odd-odd nuclei \cite{Khu21a,Khu20b}. Thus, the estimation of $P_{\rm{LRA}}$ is possible only in a case with the known $T^{\rm{exp.}}_{\alpha}$. As an example, $P_{\rm{LRA}}$ values for $^{241}$Fm, $^{250}$No, and $^{253}$Rf isotopes cannot be estimated since their ground-state $\alpha$-decays are unknown \cite{nndc}. In the two previous approaches, this was not an issue (cf. Figs.~\ref{P_LRA_empirical}a-c). 

In the neutron-deficient nuclei, an increase of $P_{\rm{LRA}}$ is observed, which is in line with results from the two previous estimates. Contrary to the $4Z-A$ systematics, now different $P_{\rm{LRA}}$ values are predicted for $^{270}$Ds and $^{286}$Fl.
At the same time, an increase in $P_{\rm{LRA}}$ as a function of $\rm log_{10}(T^{exp.}_{\alpha})$ is smoother than the increasing trends in cases of $Z^2/A$ and $4Z-A$. However, further detailed discussion of $P_{\rm{LRA}}$ as a function of $\rm log_{10}(T^{exp.}_{\alpha})$ will not lead to significant progress \cite{Wagemans86} because of the above-mentioned limit.
 
Presently, let us discuss this well-pronounced relation between the $P_{\rm{LRA}}$ and ${\rm log}_{10}(T^{\rm{exp.}}_{\alpha})$ within the $\alpha$-cluster formation point of view. 

\section{Estimation of $P_{\rm{LRA}}$ from the $\alpha$-cluster formation point of view}
\label{present_assumtpion}

In the theoretical model proposed by C\^{a}rjan \cite{Carjan76}, the LRA emission has been suggested to occur in two steps: first, the $\alpha$-cluster is formed in the nucleus, and it emits at the moment of scission. 

According to C\^{a}rjan, the $P_{\rm{LRA}}$ is defined as follows:

\begin{equation}\label{PLRA_theo.}
	P_{\rm{LRA}} = P_{\alpha} P_{\rm{s.c.}},
\end{equation}

where $P_{\alpha}$ and $P_{s.c.}$ are the probabilities of finding an $\alpha$ particle inside the nucleus and its emission during the fission (at a scission configuration \cite{Fraenkel67}), respectively. 

In such a definition, $\alpha$ particle is assumed to be formed prior to its emission during the fission. Typically, the $\alpha$-particle formation probability at the ground state is considered. Presently, this is the commonly accepted approach for the description of the LRA emission \cite{SEROT199834,Serot2000,Santhosh15}.

The $\alpha$-particle formation probability can be obtained as follows:

\begin{equation}\label{cluster}
	P_{\alpha} = T^{\rm{theo.}}_{\alpha} /T^{\rm{exp.}}_{\alpha},
\end{equation}
where $T^{\rm{theo.}}_{\alpha}$ is the theoretical $\alpha$-decay half-life. 

Combining Eqs.~\ref{PLRA_theo.} and \ref{cluster}, and taking the logarithm of both sides results:

\begin{equation}\label{PLRA_log}
	{\rm log}_{10}P_{\rm{LRA}} = {\rm log}_{10}T^{\rm{theo.}}_{\alpha} - {\rm log}_{10}T^{\rm{exp.}}_{\alpha} + {\rm log}_{10} P_{\rm{s.c.}}.
\end{equation}

Let us analyze each term of this equation. The first term can well be described by the WKB approximation \cite{Rasmussen59}, where $\alpha$ particle with an energy of $Q_{\alpha}$ penetrates throughout a potential barrier. Accordingly, this term has a well-known dependence on the $Q_{\alpha}$ value, i.e., $ {\rm log}_{10}T^{\rm{theo.}}_{\alpha}\sim Q^{-1/2}_{\alpha}$. 

The experimental $\alpha$-decay half-lives, on the other hand, are perfectly in line with the Geiger-Nuttall rule (${\rm log}_{10}T^{\rm{exp.}}_{\alpha} \sim ZQ^{-1/2}_{\alpha}$), in which effects of both $Q_{\alpha}$ value and $\alpha$-particle formation are accounted for as the empirical parameters \cite{QI2014203,Qian_2021}. 

The last term, which is in fact the most unknown, represents the emission probability of $\alpha$ particle from a ``potential well'' at the scission configuration \cite{Nobles62}. If one assumes that the $\alpha$ particle penetrates throughout a potential well, then this term can also theoretically be estimated within the WKB approach. However, since the energy of LRA is much larger than the energy of $\alpha$ particle from the ground state and is almost independent of $Z$ and $A$ of the fissioning nucleus, one can assume that (i) this term is barely dependent on the ground-state property of the fissioning nucleus. On the other hand, it is commonly accepted that the large energy of the $\alpha$ particle is a result of its acceleration in the electric field of the two heavy fragments in the neck region. This conclusion requires that $\alpha$ particle is already out of a ``potential well'' and ready for emission. In this case, the above assumption (i) can still be valid for the third term in Eq.~\ref{PLRA_log}. Despite the speculative nature of these assumptions, let us examine the dependence of $P_{\rm LRA}$ as a function of $Q^{-1/2}_{\alpha}$ on which the first two terms in Eq.~\ref{PLRA_log} are primarily dependent.  
As a summary of the aforementioned discussions, let us propose the following relationship:

\begin{equation}\label{Present}
	{\rm log}_{10}P_{\rm{LRA}} \sim {\rm log}_{10}P_{\alpha} \sim Q^{-1/2}_{\alpha}.	
\end{equation}
Here, $Q_{\alpha}$ is a value for the ground-to-ground states $\alpha$ transition. 

\subsection{Estimation of $P_{\rm{LRA}}$ as a function of alpha-decay Q-value}

In Fig.~\ref{P_LRA_empirical}(d), the $P_{\rm LRA}$ values are shown as a function of $Q_{\alpha}^{-1/2}$. It is worth noting that in all known cases of LRA emission, the $Q_{\alpha}$ values for the ground states are known \cite{nndc}. A noticeably good relation, which in fact was expected because of the above-mentioned $P_{\rm LRA}\sim \rm log_{10}(T^{exp.}_{\alpha})$ relation (see Fig.~\ref{P_LRA_empirical}b), is seen. On the other hand, the newly suggested dependence between the $P_{\rm LRA}$ and $Q_{\alpha}^{-1/2}$ is not linear, as it was in the case of $ {\rm log}_{10}T^{\rm{exp.}}_{\alpha}$, but an exponential. 

\begin{figure*}[ht]
	\vspace*{0mm}
	\centering
	\hspace*{-2mm}
	\resizebox{0.97\textwidth}{!}{\includegraphics[angle=0]{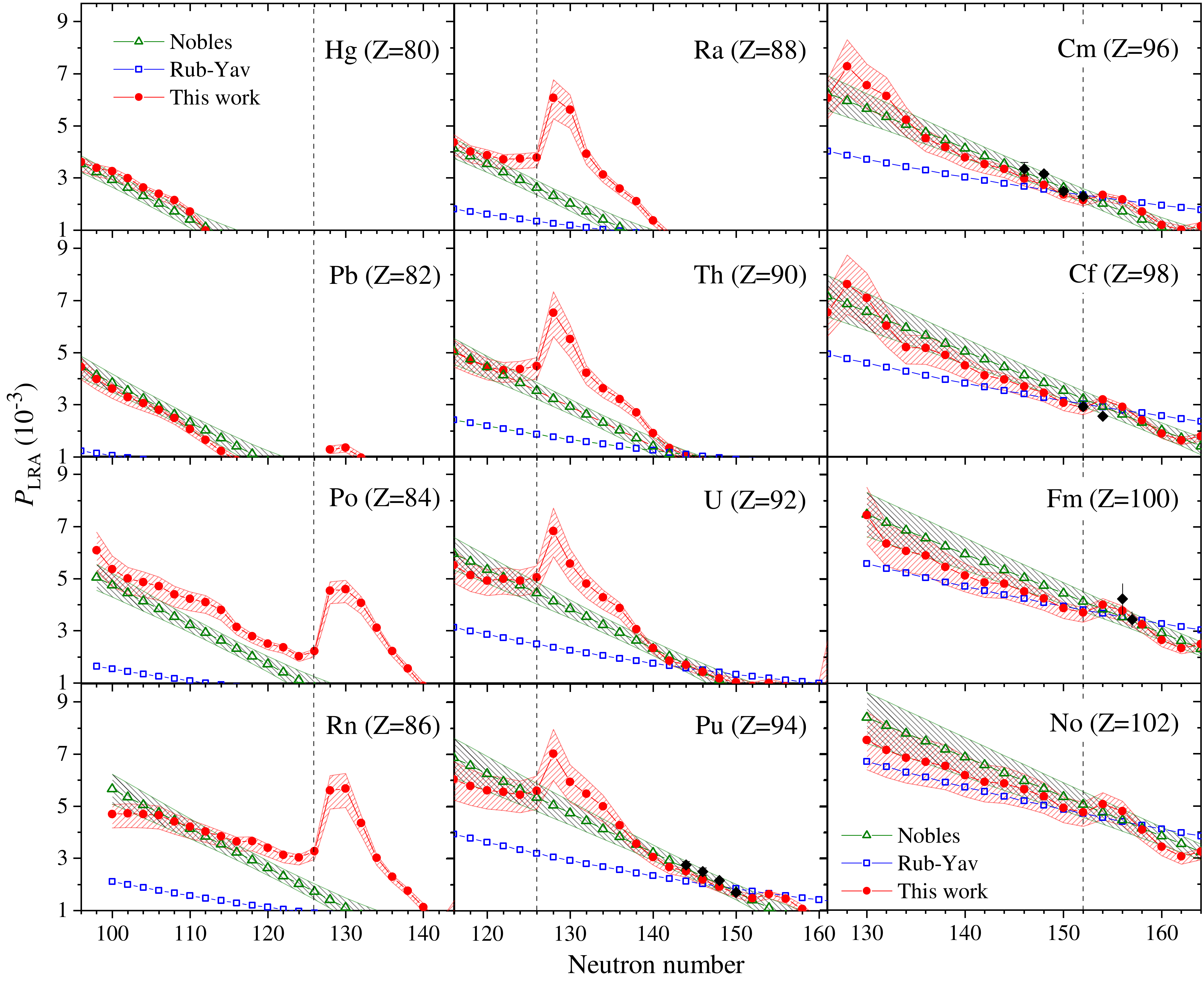}}
	\vspace*{-3mm}
	\caption{(color online) Predicted $P_{\rm{LRA}}$ values for the even-even heavy nuclei with $Z=80-102$ according to the three different approaches. Results from predictions with $Z^2/A$, $4Z-A$, and $Q^{-1/2}_{\alpha}$ variables are calculated according to the expressions given in Figs.~\ref{P_LRA_empirical}(a), (b), and (d), and are noted as Rub-Yav (Rubchenya-Yavshits) (\cite{Rubchenya88}), Nobles (\cite{Nobles62}), and this work, respectively. In the present work, the $Q_{\alpha}$ values were taken from a pure theoretical calculation \cite{Moll95}. The results from the Rubchenya-Yavshits were downscaled by a factor of 0.9 to express only the LRA emission from the total ternary fission. The experimental values are shown by the full-filled symbols. The sparse regions show boundaries of fits within the 68\% prediction level. See the text for details.}
	\label{heavy}
\end{figure*}

\begin{figure*}[ht]
	\vspace*{0mm}
	\centering
	\hspace*{-2mm}
	\resizebox{0.97\textwidth}{!}{\includegraphics[angle=0]{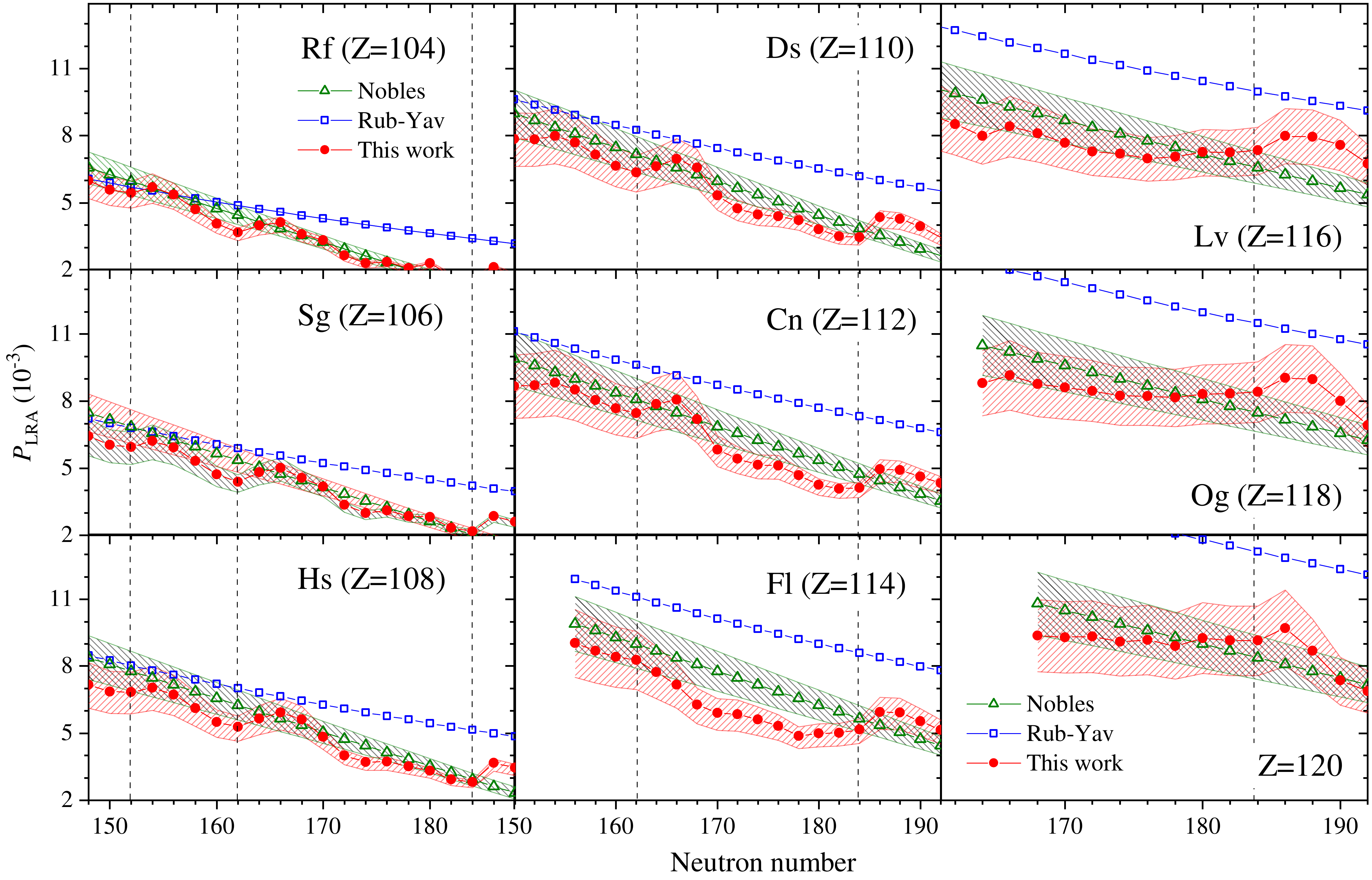}}
	\vspace*{-3mm}
	\caption{(color online) Predicted $P_{\rm{LRA}}$ values for the even-even SHN with $Z=104-120$ according to the three different approaches. See Fig.~\ref{heavy} for legends of symbols and see text for details.}
	\label{SHN}
\end{figure*}

Indeed, the experimental data are too scarce to perform a fit with sufficiently small uncertainties. Nevertheless, within the presently available experimental data, a fit was made, and the result is shown in Fig.~\ref{P_LRA_empirical}(d). The fit-result enables the estimation of the LRA emission probability by using only the ground-state $Q_{\alpha}$ value. Moreover, it permits the use of an evaluated $Q_{\alpha}$ value (e.g., from Ref.~\cite{AME2020}) for cases where the experimental $Q_{\alpha}$ is not available, like in the cases of short-lived nuclei decaying solely by SF, such as $^{241}$Fm and $^{250}$No. Predicted $P_{\rm{LRA}}$ values for these and other isotopes, which were highlighted in the previous discussions, are shown in Fig.~\ref{P_LRA_empirical}(d). The $Q_{\alpha}$ values were taken from the evaluated atomic mass table AME2020 \cite{AME2020}. However, it is worth noting that the evaluated atomic mass table does not provide the $Q_{\alpha}$ values in sufficiently wide ranges of $Z$ and/or $N$. As a consequence, one must use the theoretically calculated masses to extract $Q_{\alpha}$. In such a case, indeed, a choice of different mass tables will result in different $P_{\rm{LRA}}$ values. 
On the other hand, this will be the case for every evaluated or predicted $Q_{\alpha}$ value because a fit was performed with the experimental $Q_{\alpha}$ values. Meanwhile, by considering the large uncertainty of a fit-result, expected variations in predicted $P_{\rm{LRA}}$ values due to the $Q_{\alpha}$ from different sources can be neglected. For instance, in the case of the well-known $^{286}$Fl isotope, by using its experimental $Q_{\alpha} \approx$10.36~MeV \cite{nndc} one can calculate the $P_{\rm{LRA}}=6.6\times10^{-3}$ according to the semi-empirical expression given in Fig.~\ref{P_LRA_empirical}(d). Now, if one takes the theoretical value of $Q_{\alpha}$=9.47~MeV \cite{Moll95}, then the $P_{\rm{LRA}}$ is $5.9\times10^{-3}$, which is similarly large as the above value despite the $\approx$1~MeV reduction in the $Q_{\alpha}$ value. Thus, by considering about 16\% fit-uncertainty (68\% prediction level, see Fig.~\ref{P_LRA_empirical}d) of each, these two results can be considered as being the same, i.e., $P_{\rm{LRA}}\approx6\times10^{-3}$.

In addition, one can notice that the $P_{\rm{LRA}}$ values of $^{250}$Cf and $^{252}$Cf isotopes are somehow in contraction to trends of other isotopes as a function of $Q^{-1/2}_{\alpha}$. In fact, such a deviation can also be seen in Fig.~\ref{P_LRA_empirical}(c), where the $\rm log_{10}(T^{exp.}_{\alpha})$ was used. This may indicate the importance of a balance between $Z$ and $N$ in the nucleus, which is indeed the cornerstone of the underlying nuclear structure. In this regard, the empirical relations  $P_{\rm{LRA}}\sim Z^2/A$ and $P_{\rm{LRA}}\sim 4Z-A$ show a direct dependency on $N$.
In the present approach, $Z$ and $N$ values were not explicitly taken into account in addition to $Q_{\alpha}$, which may be considered once more experimental data becomes available. 

The estimated $P_{LRA}$ for those Fm-Fl isotopes shown in Fig.~\ref{P_LRA_empirical}(d) are similar to values from the expressions with the $Z^2/A$ and $4Z-A$ variables. 
These results once again support the anticipated increase of $P_{\rm{LRA}}$ values in neutron-deficient heavy nuclei and SHN. In this regard, the neutron-deficient $^{241,243,244}$Fm \cite{Khu08}, $^{250}$No \cite{Kal20a,Khu22a,250No_shels}, and $^{252}$No \cite{252No_Sulig} isotopes were produced with significantly large statistics, i.e., more than hundreds of their SF events. However, the experimental technique with which fission fragments of these nuclei were measured was not suitable for measuring the LRA. Thus, these cases would be the best candidates for LRA emission, which should be searched for once an appropriate experimental technique is developed.

\section{Comparison of the predictions}

Predicted $P_{\rm{LRA}}$ values for heavy nuclei and SHN with $Z=80-102$ and $Z=104-120$ according to the above three different approaches are shown in Figs.~\ref{heavy} and \ref{SHN}, respectively. To avoid an overly dense figure, only the even-even cases are plotted. Results from predictions with $Z^2/A$, $4Z-A$, and $Q^{-1/2}_{\alpha}$ variables are calculated according to the expressions given in Figs.~\ref{P_LRA_empirical}(a), (b), and (d), and are noted as Rub-Yav (Rubchenya-Yavshits) (\cite{Rubchenya88}), Nobles (\cite{Nobles62}), and this work, respectively. In the present work, the $Q_{\alpha}$ values were taken from a pure theoretical calculation \cite{Moll95}. The results from the Rubchenya-Yavshits are downscaled by a factor of 0.9 to express only the LRA emission from the total ternary fission.

Results from all three predictions are close to each other in the cases of the known LRA emission, which is indeed due to the fit.
However, in hitherto unknown cases of LRA, their results deviate from each other. Moreover, each predicts an interesting and yet unrevealed feature of the LRA emission process. The most noticeable are the results from the present work, which are very sensitive to the nucleus's shell structure. 

It is well visible that at $N=126$, 152, 162, and 184, well-pronounced local maxima in $P_{\rm{LRA}}$ values are occurring. This is due to the increase in $Q_{\alpha}$ values above significantly large shell gaps, which are predicted in the particular theory, i.e., in Ref.~\cite{Moll95}. The other two predictions are not affected by the shell structure; thus, both show a smooth dependence on $N$. In the neutron-deficient heavy nuclei (see Fig.~\ref{heavy}), $P_{\rm{LRA}}$ from Rubchenya-Yavshits are significantly smaller than the two other cases, which are close to each other. Such a tendency between the results from Nobles and the present work is preserved in cases of SHN, while the results from Rubchenya-Yavshits again deviate. Such discrepancies observed in Rubchenya-Yavshits can be attributed to its direct correlation with the fissility parameter, as shown in Fig.~\ref{P_LRA_empirical}(a). 

Nevertheless, all three approaches are predicting an increase of $P_{\rm{LRA}}$ towards the SHN. Therefore, one can conclude that the LRA emission in the SF of the SHN should not be neglected. On the other hand, it is worth noting that the LRA emission could be dependent upon the mass distribution of the fission fragments, as discussed in Ref.~\cite{SEROT199834}. Thus, fission modes of the SHN, which are still experimentally unknown in nuclei with $Z>104$ can affect their LRA emission probability. These are important in the cases of the SHN, which are relevant for the astrophysical r-process \cite{Holmbeck23,Winnet_2023}.

\section{LRA emission from the low-energy fission}

Let us consider the $P_{\rm{LRA}}$ measured from the fission of excited nuclei produced in the thermal neutron-induced reactions. In these cases, the excitation energies of compound nuclei are not exceeding $\approx7$~MeV, which prevents the contribution of a second chance fission \cite{Khu15b,Hiro17}, thus allowing the study of the LRA emission from the compound nucleus with $Z$-target and with $N$-target plus one neutron ($N+1$). It is well known that the $P_{\rm{LRA}}$ from the low-energy fission of such compound nuclei is reduced compared to the ones measured in the cases of their SF. At the same time, kinetic energies of LRA from the low-energy fission show almost the same features as the ones from SF \cite{Coleman64,Loveland67,Wagemans86,Verboven94,VERMOTE20081}. 

\begin{figure}[t]
	\vspace*{0mm}
	\centering
	\hspace*{0mm}
	\resizebox{0.48\textwidth}{!}{\includegraphics[angle=0]{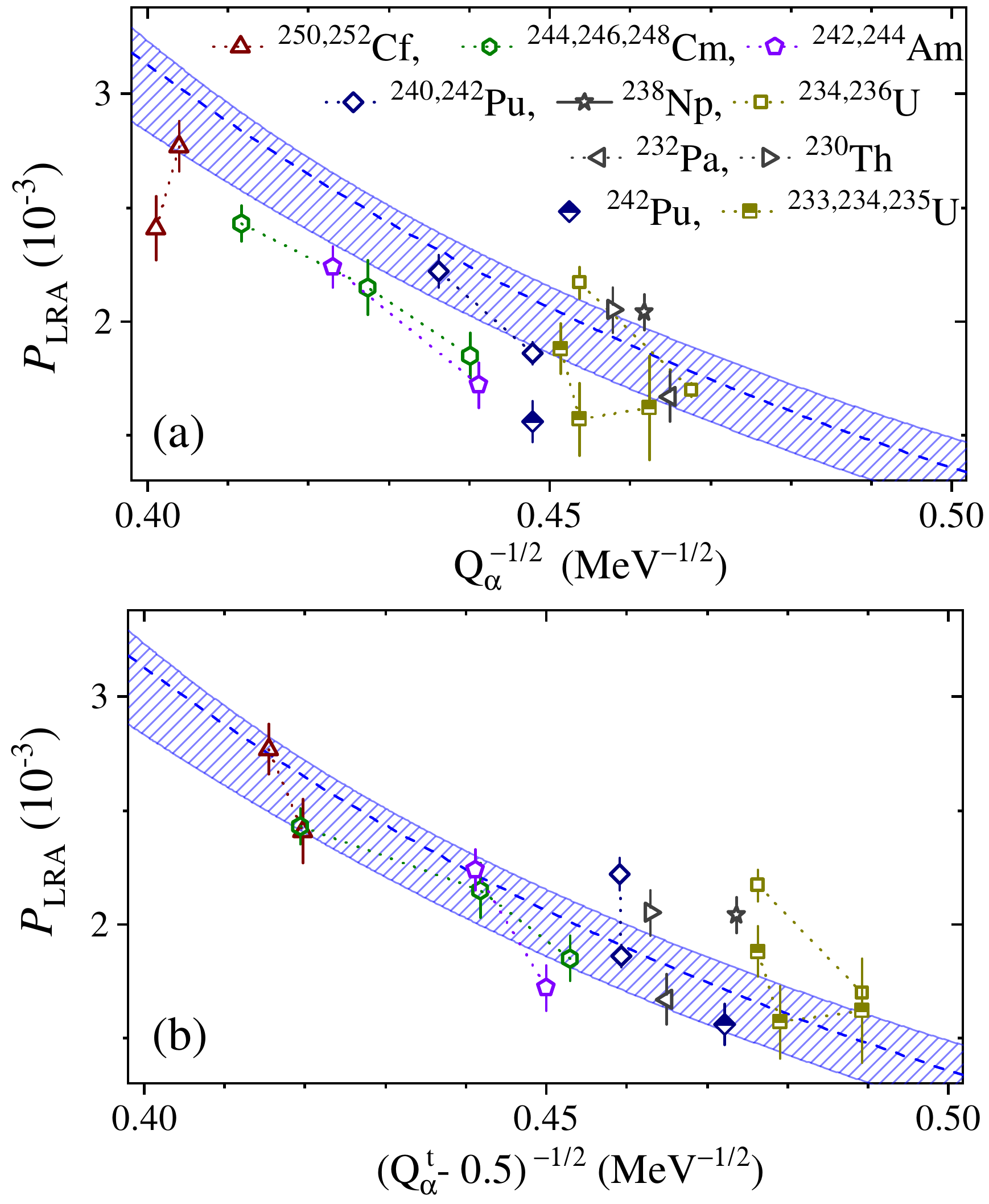}}
	\vspace*{-2mm}
	\caption{(color online) $P_{\rm{LRA}}$ measured in the cases of the excited nuclei are shown as a functions of $Q_{\alpha}$ (a). The experimental data from the thermal-neutron \cite{VERMOTE20081,WAGEMANS19811,Wagemans86,Thesis_Vermote} and photon-induced fissions \cite{Verboven94} are marked with open and half-filled symbols, respectively. The $P_{\rm{LRA}}$ values of isotopes of the same element are connected by the dotted lines. (b) The same $P_{\rm{LRA}}$ values are shown as a function the target nucleus' $Q^t_{\alpha}$ reduced by 0.5~MeV. The dashed lines and shaded regions mark the fit-result from the spontaneous fission data (see Fig.~\ref{P_LRA_empirical}(d)). The sparse regions show boundaries of fits within the 68\% prediction level. See text for details.}
	\label{predictions}
\end{figure}

In Fig.~\ref{predictions}(a), the experimental data from the thermal neutron-induced reactions, which were plotted in Fig.~\ref{PLRA_Z^2/A_exp}, are shown as a function of compound nucleus' $Q_{\alpha}$. In addition, the $P_{\rm{LRA}}$ values measured in the photon-induced fission of $^{233-235}$U and $^{242}$Pu, where the excitation energies were up to 10~MeV \cite{Verboven94}, are plotted. These energies are slightly larger than the values from thermal-neutron-induced reactions, which may be effected by the second chance fission, which is occurring from the nucleus with one neutron less. However, the second-chance-fission barriers for these nuclei are close to the initial photon energy of $\approx12$~MeV, which allows us to assume only the first-change-fission \cite{Verboven94}.

As seen in Fig.~\ref{predictions}(a), the $P_{\rm{LRA}}$ values for the low-energy fission of the heaviest nuclei are lower than the results of the fit-curve for the SF cases. However, the fitted curve and the $P_{\rm{LRA}}$ of Th, Pa, U, and Np isotopes are in line, which contradicts the commonly accepted expectation of lower values for the induced fission. On the other hand, as seen in Fig.~\ref{P_LRA_empirical}(d), only one experimental data point ($^{238}$Pu) on the SF is available in the region of $Q^{-1/2}_{\alpha}>0.45$, and it was used for the fit. Hence, it is not excluded that the yet unknown $P_{\rm{LRA}}$ values from the SF of Th, Pa, U, and Np isotopes may be larger than the ones obtained from the low-energy fission. However, the SF branches of Th, Pa, U, and Np isotopes are very small; thus, the experimental $P_{\rm{LRA}}$ values are currently unavailable. Measurements of $P_{\rm{LRA}}$ values for SF of Th-Np isotopes will be greatly helpful for both improving the fit and examining the exponential correlation, as proposed in Eq.~\ref{Present}. 

Presently, I focus on Pu-Cf isotopes, where $P_{\rm{LRA}}$ values are known from both induced fission and SF. In these isotopes, deviation from the fit curve is obvious. Let us discuss this decrease within the present concept of the ground-state $\alpha$-decay property of the fissioning nucleus. 

In the cases of excited compound nuclei produced in thermal neutron-induced reactions, a captured neutron will not form a pair, which is necessary for the formation of an $\alpha$-cluster. Accordingly, the $P_{\alpha}$ of the excited compound nucleus with $Z$ and ($N+1$) will not be the same as that of its ground-state. In this regard, a probability of finding an $\alpha$-cluster can be expressed relative to the ground-state property of the target nucleus, i.e., $Q^t_{\alpha}$. 

In addition, the excitation energy brought by the captured neutron to the compound nucleus will be shared among the various excitation modes of the nucleus. By considering an average pair-breaking energy of about 1~MeV, the amount of excitation energy brought by thermal-neutron ($\approx$7~MeV) is large enough to break a pair from the ground state. This will then further reduce the $\alpha$-cluster formation probability, i.e., $Q^t_{\alpha}$ will not be sufficient to take into account a pair-breaking effect for the emission of LRA.
In the present approach, such an effect can be expressed in a reduced $Q^t_{\alpha}$ value, i.e., by introducing an energy correction factor. In nuclear physics, it is common to use an energy correction for the $Q_{\alpha}$. For example, this is done to account for the impact of unpaired nucleons on the $\alpha$-decay rates \cite{Poe80}.

Presently, with $Q^t_{\alpha}$-0.5~MeV, the $P_{\rm{LRA}}$ values from the thermal-neutron and $\gamma$-induced reactions follow the fit-curve as shown in Fig.~\ref{predictions}(b). Accordingly, within the presently available experimental data, $P_{\rm{LRA}}$ from the excited nucleus can be described with the semi-empirical expression extracted from the SF cases. This may support the presently used assumption that the LRA emission is primarily dependent upon $P_{\alpha}$ (see Section.~\ref{present_assumtpion}).

\subsubsection{Notices on the proton-, alpha- and heavy-ion induced fission at moderate excitation energies}

Another noticeable feature of LRA from the excited nucleus is the almost independence of $P_{\rm{LRA}}$ from the further increase of excitation energy \cite{Verboven94,Coleman64,Loveland67}. For instance, for the $^{242}$Pu isotope produced in the photon \cite{Verboven94}, thermal-neutron \cite{SEROT199834}, and $\alpha$-induced \cite{Loveland67} reactions with excitation energies in the range of 6-37~MeV, an average $P_{\rm{LRA}}$ values of 1.8(2)$\times 10^{-3}$, 1.86(5)$\times 10^{-3}$, and 2.0(3)$\times 10^{-3}$ were measured, respectively. Accordingly, the above suggested estimate of $P_{\rm{LRA}}$ for the excited (up to 10~MeV) nuclei can be applied in cases of moderate excitation energies. Furthermore, it may also be suitable for cases where the compound nucleus is produced in charged-particle-induced reactions.

In this regard, let us recall an interesting result on the LRA emission from $^{213}$At \cite{Loveland67}. The experimental value for the $^{213}$At at 12~MeV excitation energy was measured 55 years ago in the $\alpha$(43~MeV)+$^{209}$Bi reaction. A relatively small value of about 6$\times 10^{-4}$ was measured. Such a small value is in line with predictions from Rubchenya-Yavshits (7$\times 10^{-4}$, Fig.~\ref{P_LRA_empirical}a) and from Nobles (1$\times 10^{-3}$, Fig.~\ref{P_LRA_empirical}b), where the nuclear shell structure was not accounted for. 

Actually, $^{213}$At is a perfect example of the nucleus having an $\alpha$ cluster, which is ``ready'' for the emission, and $^{209}$Bi as a core. The ground-state of $^{213}$At has a very large $Q_{\alpha}$ value and a very short half-life of 1.25$\times 10^{-7}$~s \cite{nndc}. According to the semi-empirical expression between the $P_{\rm{LRA}}$ and ${\rm log}_{10}(T^{\rm{exp.}}_{\alpha})$ given in Fig.~\ref{P_LRA_empirical}(c), such a short half-life results in $P_{\rm{LRA}} \approx 6\times 10^{-3}$, which is about one order of magnitude larger than the above-mentioned experimental result. Thus, the empirical relation between the $P_{\rm{LRA}}$ and ${\rm log}_{10}(T^{\rm{exp.}}_{\alpha})$ is not capable of explaining this result. 

On the other hand, as discussed above, the ground-state of $^{213}$At, which contains a ready $\alpha$ cluster, should not be considered when discussing the LRA emission from the excited compound nucleus $^{213}$At$^{\ast}$. One has to take the $\alpha$-decay property of the target nucleus $^{209}$Bi, i.e., $Q^t_{\alpha}$ of 3.14~MeV \cite{nndc}. Accordingly, by using the fit given in Fig.~\ref{P_LRA_empirical}(d) with $Q^t_{\alpha}$-0.5~MeV, one gets the $P_{\rm{LRA}}$ value of 5$\times 10^{-4}$. This is in fine agreement with the experimental data, which supports the present approach for estimating $P_{\rm{LRA}}$ from the induced fission data. 

In fact, these and similar results would greatly assist in examining the current estimations. One example is the low-fissility nucleus, which can be accessed in a beta-delayed fission. Probabilities for beta-delayed fissions have recently been calculated for the nuclei with wide ranges of $Z$ and $N$ including the SHN \cite{Khu19d}. In this regard, let us consider nuclei with much lower $Z$ and $N$ values compared to the ones discussed previously, such as $^{180}$Hg. As shown in Fig~\ref{heavy}, both Nobles and the present work agree that the $P_{\rm{LRA}}$ value for this nucleus's SF is $3\times 10^{-3}$. However, experimentally, measuring SF from this nucleus' ground state is impossible. On the other hand, the well-known beta-delayed fission from $^{180}$Tl enables the low-energy fission from $^{180}$Hg at excitation energies up to 15~MeV \cite{Andreyev10}.
In this case, one can consider $^{179}$Hg as a target nucleus, which will result in $Q^t_{\alpha}$-0.5~MeV$\approx$5.85~MeV. Based on the semi-empirical expression given in Fig.~\ref{P_LRA_empirical}(d), one calculates the $P_{\rm{LRA}}=2.8\times 10^{-3}$, which can be measured experimentally. 

Another interesting case to explore is the occurrence of LRA emission during the fission of the compound nucleus formed in the fusion reaction. In fact, all known heaviest nuclei were produced exactly in the heavy-ion induced fusion reaction, where the formed compound nucleus fissions strongly. Such studies have been conducted, and observations of relatively large $P_{\rm{LRA}}$ were reported \cite{Schad1984,Sowinski1986}. Note that these experimental results still need to be confirmed and reexamined. Nevertheless, as an example, for the $^{16}$O+$^{232}$Th reaction, $P_{\rm{LRA}}=5\times 10^{-3}$ was reported. The present semi-empirical estimate gives $P_{\rm{LRA}}=1\times 10^{-3}$, which is smaller than the above experimental value but still within the same order of magnitude. Let us consider the $^{48}$Ca+$^{209}$Bi and $^{48}$Ca+$^{249}$Cf reactions, which are leading to formations of $^{257}$Lr$^{\ast}$ and $^{297}$Og$^{\ast}$ as compound nuclei, respectively. Their corresponding $Q^t_{\alpha}$-0.5~MeV$\approx$2.64~MeV and 5.80~MeV are resulting in $P_{\rm{LRA}}$ values of 5$\times 10^{-4}$ and 3$\times 10^{-3}$ for the fission from $^{257}$Lr$^{\ast}$ and $^{297}$Og$^{\ast}$, respectively. These indicate that the LRA emission can take place during the fission of the compound nucleus formed in the heavy-ion induced reactions. On the other hand, as mentioned above, fissions from the compound nucleus are the superposition of multi-chance fissions \cite{Khu15b,Hiro17}. 
Consequently, the LRA can be emitted at every stage of the multi-chance fission process. Finally, these findings suggest that the LRA can be detected in all directions around the target. In this regard, ongoing progress on the measurements of ternary events from the multi-nucleon transfer reaction \cite{Hiro17,JEUNG2023137641,Kozulin24} may enable the study of the LRA emission from the fusion-fission and quasi-fission. 

During the experiments on synthesizing the heaviest nuclei, these LRA particles may eventually pass through the forward-angle in-flight separator and become detectable in the detection system. In this case, the LRA can be considered one of the reasons for the detection of beam-related high-energy $\alpha$ particles at the focal plane detector \cite{GatD11a,Cox_2020}.

\section{Summary, conclusion and outlook}

In the present work, the experimental probabilities of the LRA emission from spontaneous, thermal-neutron, and photon induced fissions were examined for empirical relations with various variables. The well-known correlations of LRA-emission probabilities with fissility $Z^2/A$, $4Z-A$, and ${\rm log}_{10}(T^{\rm{exp.}}_{\alpha})$ variables were discussed. In addition, on the basis of the latter correlation, i.e., $P_{\rm{LRA}}\sim {\rm log}_{10}(T^{\rm{exp.}}_{\alpha})$, and the $\alpha$ cluster point of view, a new approach for the estimation of the LRA-emission probability was proposed. This results in a semi-empirical expression, where the LRA-emission probability is described as a function of the $\alpha$-decay $Q$ value of the ground-state of fissioning nucleus.

All approaches predict substantially large $P_{\rm{LRA}}$ values up to 1\% in the known neutron-deficient heavy nuclei and SHN. This, led us to conclude that the LRA emission should not be neglected in cases of the SF of SHN. For instance, presently, the neutron-deficient Fm, No, and Rf isotopes can be produced in quantities up to thousands of nuclei, among which the LRA emission can be studied.
Moreover, new and high-intensity heavy-ion accelerator facilities, such as the SHE-Factory at FLNR, JINR, Dubna, Russia~\cite{SHE-Factory,Oga22b} and the China Accelerator Facility for Superheavy Elements (CAFE2) of the Institute of Modern Physics (IMP) in Lanzhou, China~\cite{XU2023168113}, and the planned cw-linear accelerator HELIAC at GSI Helmholtzzentrum f\"ur Schwerionenforschung, Darmstadt, Germany~\cite{Barth_2024} are capable of producing more neutron-deficient heavy nuclei and SHN. 
According to all three approaches, in the well-known $^{286}$Fl isotope, the ternary fission with the emission of LRA is predicted to occur in every about one hundred of its spontaneous fission decays. 

Finally, a crucial issue for elucidating the LRA emission from fission in the heaviest nuclei is the lack of a detection technique that is capable of identifying the three fragments simultaneously and unambiguously.

\bibliography{main_LRA}

\end{document}